# TITLE PAGE

**Title:**

Deep learning radiomics for assessment of gastroesophageal varices in people with compensated advanced chronic liver disease

**Authors and affiliations:**

Lan Wang[1]#, Ruiling He[2,3]#, Lili Zhao[4]#, Jia Wang[5]#, Zhengzi Geng[6]#, Tao Ren[7]#, Guo Zhang[8]#, Peng Zhang[1], Kaiqiang Tang[9], Chaofei Gao[1], Fei Chen[10], Liting Zhang[11], Yonghe Zhou[12], Xin Li[12], Fanbin He[6], Hui Huan[7], Wenjuan Wang[8], Yunxiao Liang[8], Juan Tang[13], Fang Ai[14], Tingyu Wang[15], Liyun Zheng[16], Zhongwei Zhao[16], Jiansong Ji[16], Wei Liu[17], Jiaojiao Xu[17], Bo Liu[17], Xuemei Wang[18], Yao Zhang[18], Qiong Yan[19], Muhan Lv[19], Xiaomei Chen[20], Shuhua Zhang[21], Yihua Wang[21], Yang Liu[21], Li Yin[22], Yanni Liu[22], Yanqing Huang[22], Yunfang Liu[23], Kun Wang[23], Meiqin Su[24], Li Bian[25], Ping An[25], Xin Zhang[25], Linxue Qian[5]*, Shao Li[1]*, Xiaolong Qi[26] *

1. Institute for TCM-X, MOE Key Laboratory of Bioinformatics, Bioinformatics Division, BNRIST, Department of Automation, Tsinghua University, Beijing, China
2. Department of Ultrasound, Donggang Branch the First Hospital of Lanzhou University, Lanzhou, China
3. The First Clinical Medical College of Lanzhou University, Lanzhou, China
4. Department of Gastroenterology and Hepatology, Tianjin Second People's Hospital, Tianjin, China

5. Department of Ultrasound, Beijing Friendship Hospital, Capital Medical University, Beijing, China

6. Department of Ultrasound, The People's Hospital of Linxia, Linxia, China

7. Department of Gastroenterology, Hospital of Chengdu Office of People's Government of Tibetan autonomous Region, Sichuan, China

8. Department of Gastroenterology, The People's Hospital of Guangxi Zhuang Autonomous Region, Nanning, China

9. Department of Control and Systems Engineering, Nanjing University, Nanjing, China

10. Department of Ultrasound, The First Hospital of Lanzhou University, Lanzhou, China

11. Department of Infectious Diseases, The First Hospital of Lanzhou University, Lanzhou, China

12. Department of Ultrasonography, Tianjin Second People's Hospital, Tianjin, China

13. Department of Infectious Disease, Zigong First people's Hospital, Zigong, China

14. Department of Ultrasound, Zigong First people's Hospital, Zigong, China

15. Department of Gastroscopy, Zigong First people's Hospital, Zigong, China

16. Department of radiology, Key Laboratory of Imaging Diagnosis and Minimally Invasive Intervention Research, Fifth Affiliated Hospital of Wenzhou Medical University, Li Shui, China

17. Ultrasound Diagnosis Center, Shaanxi Provincial People's Hospital, Xi'an, China

18. Department of Ultrasound, Ditan Hospital, Capital Medical University, Beijing, China

19. Department of Gastroenterology, The Affiliated Hospital of Southwest Medical University, Luzhou, China

20. Department of Ultrasound, The Affiliated Hospital of Southwest Medical University, Luzhou, China

21. Department of Ultrasound, North China University of Science and Technology Affiliated Hospital, Tangshan, China

22. Department of Ultrasound, Jincheng People's Hospital, Jincheng, China

23. Department of Hepatology, The Third people's Hospital of Taiyuan, Taiyuan, China

24. Department of Ultrasound, The Third people's Hospital of Taiyuan, Taiyuan, China

25. Department of Hepatology, The Sixth People's Hospital of Shenyang, Shenyang, China

26. Center of Portal Hypertension, Department of Radiology, Zhongda Hospital, Medical School, Southeast University, Nanjing, China

# These authors contributed equally to this work.

**Corresponding author:**

Xiaolong Qi, MD

Chair, Chinese Portal Hypertension Alliance (CHESS)

Chief, Center of Portal Hypertension, Department of Radiology, Zhongda Hospital,

Medical School, Southeast University, Nanjing, China

Email: qixiaolong@vip.163.com

AND

Shao Li, MD

Institute for TCM-X, MOE Key Laboratory of Bioinformatics, Bioinformatics Division, BNRIST, Department of Automation, Tsinghua University, Beijing, China

Email: shaoli@mail.tsinghua.edu.cn

AND

Linxue Qian, MD

Department of Ultrasound, Beijing Friendship Hospital, Capital Medical University, Beijing, China

Email: qianlinxue2002@163.com

**Financial Support:**

Tianjin Science and Technology Plan Project (19ZXDBSY00030); Gansu Science Fund for Distinguished Young Scholars (20JR10RA713); National Natural Science Foundation of China (62061160369).

**Compliance with Ethical Requirements**

(1) Conflict of Interest (CoI) statements

Lan Wang, Ruiling He, Lili Zhao, Jia wang, Zhengzi Geng, Tao Ren, Guo Zhang, Peng Zhang, Kaiqiang Tang, Chaofei Gao, Fei Chen, Liting Zhang, Yonghe Zhou, Xin

Li, Fanbin He, Hui Huan, Wenjuan Wang, Yunxiao Liang, Juan Tang, Fang Ai, Tingyu Wang, Liyun Zheng, Zhongwei Zhao, Jiansong Ji, Wei Liu, Jiaojiao Xu, Bo Liu, Xuemei Wang, Yao Zhang, Qiong Yan, Muhan Lv, Xiaomei Chen, Shuhua Zhang, Yihua Wang, Yang Liu, Li Yin, Yanni Liu, Yanqing Huang, Yunfang Liu, Kun Wang, Meiqin Su, Li Bian, Ping An, Xin Zhang, Linxue Qian, Shao Li, and Xiaolong Qi declared no conflicts of interest related to this study.

(2) Informed Consent in Studies with Human Subjects

All procedures followed were in accordance with the ethical standards of the responsible committee on human experimentation (institutional and national) and with the Helsinki Declaration of 1975, as revised in 2008 (5). Informed consent was obtained from all patients for being included in the study.

**Author contributions**

Study conception and design: Xiaolong Qi, Shao Li, Lan Wang and Ruiling He;

Data collection and analysis: Ruiling He, Lan Wang, Lili Zhao, Jia Wang, Zhengzi Geng, Tao Ren, Guo Zhang, Peng Zhang, Kaiqiang Tang, Chaofei Gao, Fei Chen, Liting Zhang, Yonghe Zhou, Xin Li, Fanbin He, Hui Huan, Wenjuan Wang, Yunxiao Liang, Juan Tang, Fang Ai, Tingyu Wang, Liyun Zheng, Zhongwei Zhao, Jiansong Ji, Wei Liu, Jiaojiao Xu, Bo Liu, Xuemei Wang, Yao Zhang, Qiong Yan, Muhan Lv, Xiaomei Chen, Shuhua Zhang, Yihua Wang, Yang Liu, Li Yin, Yanni Liu, Yanqing Huang, Yunfang Liu, Kun Wang, Meiqin Su, Li Bian, Ping An, Xin Zhang;

The first draft of the manuscript: Lan Wang, Ruiling He;

**ABSTRACT**

**Objective:** Bleeding from gastroesophageal varices (GEV) is a medical emergency associated with high mortality. We aim to construct an artificial intelligence-based model of two-dimensional shear wave elastography (2D-SWE) of the liver and spleen to precisely assess the risk of GEV and high-risk gastroesophageal varices (HRV).

**Design:** A prospective multicenter study was conducted in patients with compensated advanced chronic liver disease. 305 patients were enrolled from 12 hospitals, and finally 265 patients were included, with 1136 liver stiffness measurement (LSM) images and 1042 spleen stiffness measurement (SSM) images generated by 2D-SWE. We leveraged deep learning methods to uncover associations between image features and patient risk, and thus conducted models to predict GEV and HRV.

**Results:** A multi-modality Deep Learning Risk Prediction model (DLRP) was constructed to assess GEV and HRV, based on LSM and SSM images, and clinical information. Validation analysis revealed that the AUCs of DLRP were 0.91 for GEV (95% CI 0.90 to 0.93, $p < 0.05$) and 0.88 for HRV (95% CI 0.86 to 0.89, $p < 0.01$), which were significantly and robustly better than canonical risk indicators, including the value of LSM and SSM. Moreover, DLPR was better than the model using individual parameters, including LSM and SSM images. In HRV prediction, the 2D-SWE images of SSM outperform LSM ($p < 0.01$).

**Conclusion:** DLRP shows excellent performance in predicting GEV and HRV over canonical risk indicators LSM and SSM. Additionally, the 2D-SWE images of SSM provided more information for better accuracy in predicting HRV than the LSM.

**Introduction**

The gastroesophageal varices (GEV) is a common and severe complication of compensated advanced chronic liver disease (cACLD).[1-3] Variceal hemorrhage is associated with a six-week mortality rate of between 15% and 25%.[2,3] Thus, the accurate risk assessment of variceal hemorrhage is essential for the prognosis, surveillance and management of patients with cACLD, especially in detecting high-risk gastroesophageal varices (HRV) and determining the need for primary preventative therapy.[2-4]

Esophagogastroduodenoscopy (EGD) is considered the gold standard for risk assessment of varices in patients with cACLD.[2,3] However, it is invasive, poorly tolerated and limited by various potential complications.[5] Consequently, patients with cACLD may decline a screening procedure, because they are stable and asymptomatic. The combined model based on liver stiffness measurement (LSM) by transient elastography (TE) and platelet count (PLT) has been strongly recommended by the Baveno VI consensus.[6] However, the potential limitation is that the scale of patients meeting the Baveno VI standard is relatively low.[7,8] Previous studies have found that spleen stiffness measurement (SSM) combined with Baveno VI by TE can significantly improve the proportion of patients who safely avoid endoscopic screening.[9,10] However, the maximum threshold for measuring tissue stiffness by a standard TE probe may not be sufficient to evaluate spleen stiffness.[11] A novel spleen-dedicated probe, which avoids the above limitations, has recently been commercialized but has yet to be widely used.[4,12]

With the development of two-dimensional shear wave elastography (2D-SWE) imaging technology and its wide application in the clinic, researchers can obtain a large number of reliable images of liver and spleen stiffness.[13,14] A current methodological bottleneck is how to analyze based on these complex image data intelligently. The artificial intelligence algorithm developed in recent years can extract useful features from complex data.[15-17] In this regard, deep learning is widely used in radiomics. It has the potential to uncover disease characteristics that fail to be appreciated by the naked eye.[18] Deep learning for radiomics can use more valuable information from images, which has been proven helpful in assessing liver fibrosis, where liver images based on 2D-SWE were acquired for analysis.[19] Recently, research has reported that the performance of the deep learning model is improved with the fusion approach that integrates image features and clinical information.[20]

To move beyond the limitations of previous radiomics approaches and accelerate the broader adoption of deep learning technology based on radiomics by clinicians, in this study, we have developed the multi-modality deep learning risk prediction model (DLRP), a new tool for assessing GEV and HRV in patients with cACLD by 2D-SWE images (including: liver elastography images (LE), liver grayscale images (LG), spleen elastography images (SE), spleen grayscale images (SG)) and clinical information. We aimed to develop a deep learning tool to non-invasively and accurately evaluate GEV and HRV in cACLD patients.

**Materials and methods**

*Design and overview*

This was a multicenter, prospective study. Patients with cACLD who volunteered to participate in this study were recruited from 12 hospitals in different regions of China from October 2020 to September 2022. This multicenter study was approved by the ethics committee of the principal investigator's hospital and was registered at ClinicalTrials.gov (NCT04546360).

*Patient enrolments*

Three hundred five potentially eligible patients from 12 hospitals were prospectively enrolled in this study. The inclusion criteria were: 1) age older than 18 years; 2) confirmed cACLD whose value of LSM by 2D-SWE ≥ 8 kPa;[21] 3) without decompensated events (e.g., ascites, bleeding, or overt encephalopathy) in the past; 4) with LSM and SSM by 2D-SWE, laboratory tests and EGD; 5) with written informed consent. Patients were excluded for the following: 1) unqualified acquisitions for LSM or SSM; 2) the time frame between 2D-SWE, laboratory data and EGD>1 month; 3) accepted primary prevention (non-selective beta blockers or endoscopic variceal ligation); 4) splenectomy, splenic embolism or absence of spleen; 5) hepatocellular carcinoma. A detailed flow chart pertinent to the study design was illustrated in Figure 1.

*Two-dimensional shear wave elastography*

All patients underwent LSM and SSM based on 2D-SWE in fasting patients (fasting for more than six hours). The Aixplorer ultrasound system (Supersonic Imagine, SSI, France) with an abdominal 3.5 MHz curved array probe was used. The operator was blind to the patients' EGD findings.

LSM was measured in the right lobe using the 7th to 9th rib intercostal approach, with the right arm in maximum abduction. The region of interest, size 4 cm×3 cm and fan-shaped, was placed in an area of parenchyma free of large vessels and bile ducts, avoiding noisy areas from rib shadowing. Patients should hold their breath for a few seconds when a suitable image is obtained. Then, the 2D-SWE measurement was started, and the Q-Box was placed at least 1 cm and no more than 3 cm away from the liver capsule, and the diameter was not less than 1.5 cm. At the same time, the liver images were saved (figure 2A). The latest EFSUMB guidelines were followed, including the reliability of LSM during the protocol of performing LSM in this study.[15]

SSM was performed in the right lateral decubitus position with the left arm raised on the head to expose intercostal space fully. The probe was placed in the 9th to 11th intercostal space of the posterior axillary line on the left. The size of the region of interest was not less than 3 cm×2 cm and was placed between the parenchyma of the central region and the lower pole without large blood vessels or abnormal lesions. The Q-Box was placed at least 1 cm and no more than 3 cm away from the spleen capsule, and the diameter was not less than 1.0 cm. Other steps are the same as above. At the same time, the spleen images were saved (figure 2B).

In each case, the stiffness of the liver and spleen was measured at least three times. The value of LSM and SSM were depicted in kilopascals (kPa). The reliable value of LSM was defined as the stability index of image ≥ 80% and interquartile range/median ratio < 30%. Although the published guidelines or expert consensus do not unify the standard for the reliable value of SSM, it is considered that the uniform color filling in the region of interest and more than 2/3 of color filling are effective measurements in our study, regardless of the stability index and interquartile range/median ratio which have been developed on the standard of LSM.

In image acquisition, synchronously measure and record spleen diameter, spleen thickness, splenic vein and portal vein diameter by the operator of liver and spleen stiffness measurement.

*Endoscopic evaluation*

Two independent gastroenterology experts performed EGD (>1000 endoscopies each of them) in all patients. A flexible EVIS EXERA video gastroscope (Olympus Europa Medical Systems, Hamburg, Germany) was used. HRV was defined as large varices (varix size ≥ 5 mm) or small varices (varix size < 5 mm) combined with red wale or Child-Puch C.[4] The procedure was carried out up to 1 month after or before 2D-SWE, and operators had no information about the results of 2D-SWE.

*Laboratory examinations*

Laboratory examinations, including serum aspartate aminotransferase (AST), alanine aminotransferase (ALT), Glutamyl transpeptidase (GGT), alkaline phosphatase (ALP), bilirubin, serum albumin, serum creatinine, platelet count, prothrombin time and international normalized ratio (INR) were measured. The severity of liver disease was determined by the Child-Pugh score and the MELD score was calculated according to the UNOS formula.

*Deep Learning Risk Prediction model of Single-modality (DLRP-SM)*

In this section, we propose a new diagnostic approach based on single-modality deep learning named DLRP-SM, using 2D-SWE images (e.g., LE, LG, SE and SG) to predict GEV and HRV.

In order to enable DLRP-SM to extract the features of the effective area more effectively and avoid the influence of irrelevant edge noise information, our preprocessing methods for 2D-SWE images are as follows: 1) Image segmentation of elastography images and grayscale images; 2) Determining the target position Q-box and radius with Hough circle detection; 3) Cropping a square area as the deep learning research area, with the center of mass as the center of the Q-box, and the side length is four times the radius of the Q-box.

All patients were randomly divided into training and verification cohorts. The training cohort was used to train the designed deep neural network whose parameters were optimized by the Adam optimizer, and the test cohort was used to evaluate the performance of the trained model. To extract the features of the medical images, we used the framework of Resnet, whose parameters were pre-trained on the ImageNet in advance.[22] Conventional CNN networks suffer from vanishing gradient and network degradation when the number of network layers is deepened. In contrast to regular CNN networks, Resnet adds a direct channel to the network, allowing it to retain a certain proportion of the output of the previous network layer. At the same time, the training accuracy is improved and computational complexity is reduced. Compared with ImageNet, one of the world's largest visual databases for image recognition, our medical dataset has less quantity and higher similarity, which requires more abundant features for classification. Therefore, we fine-tuned the Resnet parameters pre-trained in the ImageNet dataset with ultrasound images consisting of elastography images and grayscale images. To prevent the model from overfitting, we randomly cut, flipped and rotated all image data. We passed them through the Resnet with the Bottleneck and residual units in training. The image features are obtained after passing through 12 layers of Bottleneck with an adaptive average pooling operator, and then flattened to 2048 *1. The final classification results are output by the softmax layer. In the single-modality experiment, we trained two binary classification networks respectively to judge whether there is GEV or HRV.

*DLRP*

Medical data is often multimodal. For example, elastography images, grayscale images and clinical data all contained relative information about this disease. Clinical data consists of routine ultrasound results, endoscopic evaluation results and laboratory examination results. Thus, in this section, we divided them into image and clinical data modalities.

In order to utilize the clinical data more effectively, our preprocessing method is as follows: 1) We separated the numerical features and textual features, where the numerical features were normalized between 0 and 1 by min-max normalization, and the textual features were encoded into numeric vectors based on keywords. 2) Using the Pearson correlation coefficient, the top 60% of the data with the correlation coefficient was taken as the clinical information modality input to the model.

As shown in Figure 2C, for each type of images (LE, LG, SE and SG), we passed them through the Bottleneck of the Resnet to extract the relative features. Then, the four types of image features are compressed into feature vectors of different lengths through the gating mechanism, and the final image feature vectors are obtained by concatenating.

Next, we use the feature embedding method to align the feature vectors of clinical data and medical images to use the patient information comprehensively. The clinical data is input into a Multilayer Perceptron after feature filtering and mapping to obtain corresponding feature vectors. The final classification results are output by the softmax layer. In the multi-modality experiment, similar to the single-modality part, we trained two binary classification networks to judge whether there is GEV or HRV, and we adopted the Dropout method to discard a certain proportion of model parameters.

*Diagnostic robustness evaluation of DLRP*

Twelve Chinese hospitals participated in this study. The patient's image data and clinical data were collected from each hospital. Due to the difference of operators in different centers, images have significant heterogeneity (e.g., the size, depth, gain). Therefore, we performed image clipping, location of interest extraction, image gain adjustment and other operations to improve the robustness of the model. In the experiment, all the patients were randomly selected from these hospitals. About 70% were formed into training cohorts, and the remaining patients were formed into validation cohorts. Finally, we used the above data to validate the risk of GEV and HRV in patients with cACLD.

*Statistical analysis*

The prediction results were validated by quantitative indexes, including sensitivity, specificity, positive predictive value and negative predictive value. The Chi-squared test and t test were used to determine whether there was any statistical difference in patient characteristics. The area under the receiver operating characteristic (ROC) curve (AUC) was used to estimate the correct prediction probability of the model. Delong's test was used to test whether there is a statistical difference in risk prediction between DLRP and other methods.

**Results**

*Patient characteristics*

A total of 265 patients with cACLD were finally enrolled for analysis between October 2020 and September 2022 from 12 medical centers in China, with 1136 liver images and 1042 spleen images by 2D-SWE (figure 1). Among them, GEV was documented in 142 (53.6%) and HRV in 83 (31.3%) patients. After randomization of these patients, 191 patients with 808 liver images and 187 patients with 742 spleen images were assigned to the training cohort. The other 74 patients with 328 liver images and 72 patients with 300 spleen images composed the validation cohort.

The baseline characteristics of the study population are summarized in Table 1. Overall, the primary etiology (213 [80.4%]) of cACLD was hepatitis B infection, and the median (IQR) value of LSM and SSM were 14.0 (8.2) kPa and 32.9 (15.2) kPa. Between the training and validation cohorts, there were neither significant differences in all baseline characters (p>0.05), nor the distribution of patients in GEV/Non-GEV and HRV/Non-HRV(p>0.05).

***DLRP-SM***

1) Classifying whether it is GEV with 2D-SWE images

For the training cohort, the training results are shown in Figure 3A. Compared with the value of LSM and SSM, DLRP-SM, including DLRP-LE, DLRP-LG, DLRP-SE and DLRP-SG, showed higher diagnostic accuracy for classifying whether it is GEV, and the differences of AUCs were statistically significant (all $p<0.001$, table 2). AUCs of DLRP-LE, DLRP-LG, DLRP-SE and DLRP-SG reached startling 0.98, 0.97, 0.99 and 0.98, which were higher than the LSM and SSM. The sensitivity and specificity analyses also demonstrated that DLRP-LE, DLRP-LG, DLRP-SE and DLRP-SG were universally better than LSM and SSM.

For the validation cohort, the validation results are shown in Figure 3C. AUCs of DLRP-LE and DLRP-SE reached 0.85 and 0.81, significantly higher than DLRP-LG, DLRP-SG, LSM and SSM for classifying whether or not it is GEV (all $p<0.01$, table 2). The sensitivity and specificity analyses also demonstrated that DLRP-LE and DLRP-SE were universally better than other methods (table 2).

2) Classifying whether it is HRV with 2D-SWE images

For the training cohort, the training results are shown in Figure 3B. Compared with the value of LSM and SSM, DLRP-SM, including DLRP-LE, DLRP-LG, DLRP-SE and DLRP-SG, showed higher diagnostic accuracy for classifying whether or not it is HRV, and differences of AUCs were statistically significant (all $p<0.001$, table 3).

AUCs of DLRP-LE, DLRP-LG, DLRP-SE and DLRP-SG reached 0.97, 0.98, 0.97 and 0.95, which were higher than the LSM and SSM. The sensitivity and specificity analyses also demonstrated that DLRP-SM methods were universally better than LSM and SSM.

For the validation cohort, the validation results are shown in Figure 3D. The AUC of DLRP-SE reached 0.83, significantly higher than DLRP-LE, DLRP-LG, DLRP-SG, LSM and SSM for classifying whether or not it is HRV (all p<0.01, table 3). The sensitivity and specificity analyses also demonstrated that DLRP-SE was universally better than other methods.

### *DLRP*

We use the combination of 2D-SWE images of the liver and spleen with clinical information to predict GEV and HRV, called a multi-modality model. In the multi-modality experiment, we choose the optimal DLRP-SM method to compare with the multi-modality method DLRP. When classification of whether it is GEV or HRV, we choose DLRP-LE and DLRP-SE as the DLRP-SM, respectively, because they perform best in the single-modality experiment. In the training cohort, DLRP demonstrated statistically higher AUCs than DLRP-SM (p < 0.001) (figure 4A, table 4). In the test cohort, the AUCs of DLRP reached 0.91 and 0.88 for diagnosing GEV and HRV, respectively, which were higher than DLRP-SM (figure 4B, all p <0.001). The sensitivity and specificity analyses also demonstrated that DLRP was universally better than DLRP-SM for assessing of GEV and HRV (table 2, table 3, table 4).

*Diagnostic robustness evaluation of DLRP*

Three randomly selected combinations of hospitals were employed to establish three different training cohorts (with the same number of patients), so the DLRP with three cohorts of parameters was obtained, respectively. In either the training or validation cohort, the results in three ROC curves always overlapped each other (figure 5), and no significant differences were found (all p>0.05, table 5). These revealed that DLRP demonstrated robust and consistent performances regardless of the data from which medical centers, as long as the number of enrolled patients in different training cohorts was fairly constant.

**Discussion**

In this multicenter prospective study, the diagnostic accuracy of DLRP was significantly improved compared with LSM and SSM in assessing GEV and HRV in patients with cACLD. In the evaluation of GEV, the AUCs of DLRP, DLRP-LE and DLRP-SE in the training cohort were 0.97, 0.98 and 0.99, respectively, and the AUCs in the validation cohort were 0.91, 0.85 and 0.81, respectively (figure 3A, figure 3C, table 2). It can be seen that DLRP has a good assessment effect on GEV. In the evaluation of HRV, the AUCs of DLRP, DLRP-LE and DLRP-SE in the training cohort are all 0.97, and the AUCs in the verification cohort are 0.88, 0.71 and 0.83, respectively, showing that DLRP has a good assessment effect on HRV (figure 3B, figure 3D, table 3). Accordingly, when further integrating other clinical information, including demographic and serological indicators, the performance was better than only using image information in both the training cohort and validation cohort (figure 4, table 2, table 3, table 4).

The above results show that DLRP can be successfully used in assessing GEV and HRV in cACLD patients, and performs better than only LSM and SSM values. And DLRP reduces the unnecessary endoscopic screening of patients with cACLD, which may be a potential breakthrough in GEV and HRV assessment.

The second finding of our study is that both DLRP-LE and DLRP-SE show good performance when evaluating GEV. In contrast, DLRP-SE performed significantly better than DLRP-LE in evaluating HRV. It reveals that 2D-SWE images of the spleen provide more information for better accuracy in assessing HRV. Traditional Chinese medicine proposes that liver disease will affect the spleen, and the spleen and stomach affect and reflect each other. Meanwhile, in these two classification experiments, DLRP-LE and DLRP-SE perform significantly better than DLRP-LG and DLRP-SG, which reveals that 2D-SWE images contain more critical information than grayscale images.

DLRP is a simplified and novel artificial intelligence tool from the perspective of images. There is a broader range of image utilization, compared with only using the values of LSM and SSM, which only uses images in the circular area of Q-Box (around 2 cm$^2$) to measure the average liver and spleen stiffness. DLRP fully uses a larger area (around 9 cm$^2$), including the Q-Box surrounding information. At the same time, regions of interest are automatically extracted. Unlike other methods that require manual labeling, we automatically identify regions of interest, reducing the workload of doctors. Furthermore, DLRP uses deep learning methods to extract image features automatically. By extracting various hidden features in images, which reflect the heterogeneity of image intensity and texture, patients are classified according to the corresponding GEV and HRV metrics. These provides a more comprehensive assessment than analysis using the value of LSM and SSM.

Finally, DLRP showed remarkable robustness in this multicenter study. When the cohorts were constructed using a randomly selected combination of patients, no significant difference was found in the classification of GEV and HRV between the training and validation cohorts using DLRP. Moreover, the evaluation performance was consistent with the DLRP (table 5). These results demonstrate that DLRP is robust, reliable and valuable for clinical promotion. Using data from a limited number of hospitals to train and build a DLRP is likely sufficient for its application to evaluating GEV and HRV in clinical practice.

To the best of our knowledge, DLRP is the first simplified and novel artificial intelligence tool from the perspective of images that can accurately and non-invasively identify GEV and HRV. Moreover, our work is the first prospective study combining radiomics, LSM and SSM based on 2D-SWE, and clinical information for assessing GEV and HRV in cACLD patients. We collected 1136 liver images and 1042 spleen images from 265 patients in 12 medical centers, and we believe it is the largest 2D-SWE radiomics assessment of GEV and HRV in cACLD patients to date. Image acquisition for each patient is subject to stringent quality controls. The final results proved that DLRP is essential in evaluating GEV and HRV in cACLD patients. With the DLRP, the operator only needs to carry out the daily workflow of 2D-SWE to automatically locate and analyze the key positions, which is very convenient for clinical applications.

In addition, we found several studies applying radiomics to 2D-SWE analysis. Gatos et al. reported a multicenter study using hard-coded radiological features extracted from 2D-SWE images to identify chronic liver disease patients from healthy individuals, and the proposed method achieved an AUC of 0.87.[23] Kun Wang et al. reported a multicenter study that analyzed the liver images generated by 2D-SWE using convolutional neural networks to predict the fibrosis stage of chronic hepatitis B patients, which was superior to the value of LSM.[19] However, their methods are fundamentally different from the present study. These methods neither provided risk assessment for GEV and HRV in the patient with cACLD nor incorporated other clinical data.

One of the limitations of our study is the limited number of patients included. In future studies, we need to involve more cACLD patients on a larger scale to train the deep learning model better. We can also improve the overall performance of DLRP by introducing an attention mechanism to self-learn the modality importance for multi-modality fusion to analyze the available data more deeply. At the same time, we can apply the image method to other related research, such as the discovery of disease mechanisms and drug efficacy monitoring.[24-26] In addition to these limitations, our study did not consider the patients such as non-alcoholic fatty liver disease. It included the cACLD patients mainly infected with the hepatitis B virus. Moreover, we did not compare the effects of 2D-SWE instruments from different manufacturers. These deserve further study in the future.

In conclusion, this study presents a non-invasive method for assessing GEV and HRV, demonstrating this method's good performance in cACLD patients. All these indicate that DLRP has a good potential for clinical promotion, and further studies in larger patient populations are needed. Among them, the 2D-SWE images of SSM provided more information for predicting HRV.

# References

1. Qi X, Berzigotti A, Cardenas A, et al. Emerging non-invasive approaches for diagnosis and monitoring of portal hypertension. The Lancet Gastroenterology & Hepatology 2018; 3(10): 708-719. DOI: 10.1016/S2468-1253(18)30232-2.

2. European Association for The Study of the Liver. EASL Clinical Practice Guidelines for the management of patients with decompensated cirrhosis. Journal of hepatology 2018; 69(2): 406-460. DOI: 10.1016/j.jhep.2018.03.024.

3. Garcia-Tsao G, Abraldes J G, Berzigotti A, et al. Portal hypertensive bleeding in cirrhosis: Risk stratification, diagnosis, and management: 2016 practice guidance by the American Association for the study of liver diseases. Hepatology 2017; 65(1): 310-335. DOI: 10.1002/hep.28906.

4. de Franchis R, Bosch J, Garcia-Tsao G, et al. Baveno VII - Renewing consensus in portal hypertension. Journal of hepatology 2022; 76(4):959-974. DOI: 10.1016/j.jhep.2021.12.022.

5. Edelson J, Suarez A L, Zhang J, et al. Sedation during endoscopy in patients with cirrhosis: safety and predictors of adverse events. Digestive Diseases and Sciences 2020; 65:1258-1265. DOI: 10.1007/s10620-019-05845-7.

6. Franchis R, Baveno VI Faculty. Expanding consensus in portal hypertension: Report of the Baveno VI Consensus Workshop: Stratifying risk and individualizing care for portal hypertension. Journal of hepatology 2015; 63(3):743-752. DOI: 10.1016/j.jhep.2015.05.022.

7. Jangouk P, Turco L, De Oliveira A D, et al. Validating, deconstructing and

refining Baveno criteria for ruling out high-risk varices in patients with compensated cirrhosis. Liver international 2017; 37(8):1177-1183. DOI: 10.1111/liv.13379.

8. Maurice J B, Brodkin E, Arnold F, et al. Validation of the Baveno VI criteria to identify low risk cirrhotic patients not requiring endoscopic surveillance for varices. Journal of hepatology 2016; 65:899-905. DOI: 10.1016/j.jhep.2016.06.021.

9. Colecchia A, Ravaioli F, Marasco G, et al. A combined model based on spleen stiffness measurement and Baveno VI criteria to rule out high-risk varices in advanced chronic liver disease. Journal of hepatology 2018; 69:308-317. DOI: 10.1016/j.jhep.2018.04.023.

10. Wang H, Wen B, Chang X, et al. Baveno VI criteria and spleen stiffness measurement rule out high-risk varices in virally suppressed HBV-related cirrhosis. Journal of hepatology 2021; 74:584-592. DOI: 10.1016/j.jhep.2020.09.034.

11. Calvaruso V, Bronte F, Conte E, et al. Modified spleen stiffness measurement by transient elastography is associated with presence of large oesophageal varices in patients with compensated hepatitis C virus cirrhosis. Journal of viral hepatitis 2013; 20:867-74. DOI: 10.1111/jvh.12114.

12. Stefanescu H, Marasco G, Calès P, et al. A novel spleen-dedicated stiffness measurement by FibroScan® improves the screening of high-risk oesophageal varices. Liver international 2020; 40: 175-185. DOI: 10.1111/liv.14228.

13. Dietrich C F, Bamber J, Berzigotti A, et al. EFSUMB Guidelines and Recommendations on the Clinical Use of Liver Ultrasound Elastography, Update 2017 (Long Version). Ultraschall in der medizin 2017; 38(4):e16-e47. DOI: 10.1055/s-0043-103952.

14. Ferraioli G, Filice C, Castera L, et al. WFUMB guidelines and recommendations for clinical use of ultrasound elastography: Part 3: liver. Ultrasound in medicine & biology 2015; 41(5):1161-1179. DOI: 10.1016/j.ultrasmedbio.2015.03.007.

15. Lei Y, Li S, Liu Z, et al. A deep-learning framework for multi-level peptide–protein interaction prediction. Nature Communications 2021; 12 (1): 1-10. DOI: 10.1038/s41467-021-25772-4.

16. Brown J R G, Mansour N M, Wang P, et al. Deep learning computer-aided polyp detection reduces adenoma miss rate: a United States multi-center randomized tandem colonoscopy study (CADeT-CS Trial). Clinical Gastroenterology and Hepatology 2022; 20(7): 1499-1507. DOI: 10.1016/j.cgh.2021.09.009.

17. Gao Y, Xin L, Lin H, et al. Machine learning-based automated sponge cytology for screening of oesophageal squamous cell carcinoma and adenocarcinoma of the oesophagogastric junction: a nationwide, multicohort, prospective study. The Lancet Gastroenterology & Hepatology 2023; 8(5): 432-445. DOI: 10.1016/S2468-1253(23)00004-3.

18. Avanzo M, Wei L, Stancanello J, et al. Machine and deep learning methods for radiomics. Medical Physics 2020; 47(5):e185-e202. DOI: 10.1002/mp.13678.

19. Wang K, Lu X, Zhou H, et al. Deep learning Radiomics of shear wave

elastography significantly improved diagnostic performance for assessing liver fibrosis in chronic hepatitis B: a prospective multicenter study. Gut 2019; 68(4): 729-741. DOI: 10.1136/gutjnl-2018-316204.

20. Joo S, Ko ES, Kwon S, et al. Multimodal deep learning models for the prediction of pathologic response to neoadjuvant chemotherapy in breast cancer. Scientific reports 2021; 11(1):18800. DOI: 10.1038/s41598-021-98408-8.

21. Herrmann E, Lédinghen V, Cassinotto C, et al. Assessment of biopsy-proven liver fibrosis by two-dimensional shear wave elastography: An individual patient data-based meta-analysis. Hepatology 2018; 67(1): 260-272. DOI: 10.1002/hep.29179.

22. He K, Zhang X, Ren S, et al. Deep residual learning for image recognition. Proceedings of the IEEE conference on computer vision and pattern recognition 2016; 770-778. DOI: 10.1109/CVPR.2016.90.

23. Gatos I, Tsantis S, Spiliopoulos S, et al. A machine-learning algorithm toward color analysis for chronic liver disease classification, employing ultrasound shear wave elastography. Ultrasound in medicine & biology 2017; 43(9): 1797-1810. DOI: 10.1016/j.ultrasmedbio.2017.05.002.

24. Lin X, Hu L, Gu J, et al. Choline kinase α mediates interactions between the epidermal growth factor receptor and mechanistic target of rapamycin complex 2 in hepatocellular carcinoma cells to promote drug resistance and xenograft tumor progression. Gastroenterology 2017; 152(5): 1187-1202. DOI: 10.1053/j.gastro.2016.12.033.

25. Zhao X, Fu J, Xu A, et al. Gankyrin drives malignant transformation of chronic

liver damage-mediated fibrosis via the Rac1/JNK pathway. Cell death & disease 2015; 6(5): e1751. DOI: 10.1038/cddis.2015.120.

26. Yuan Z, Zhou Q, Cai L, et al. SEAM is a spatial single nuclear metabolomics method for dissecting tissue microenvironment. Nature Methods 2021; 18: 1223-1232. DOI: 10.1038/s41592-021-01276-3.

**Figure 1.** Flowchart of study enrollment.
2D-SWE, two-dimensional shear wave elastography; EGD, esophagogastroduodenoscopy.

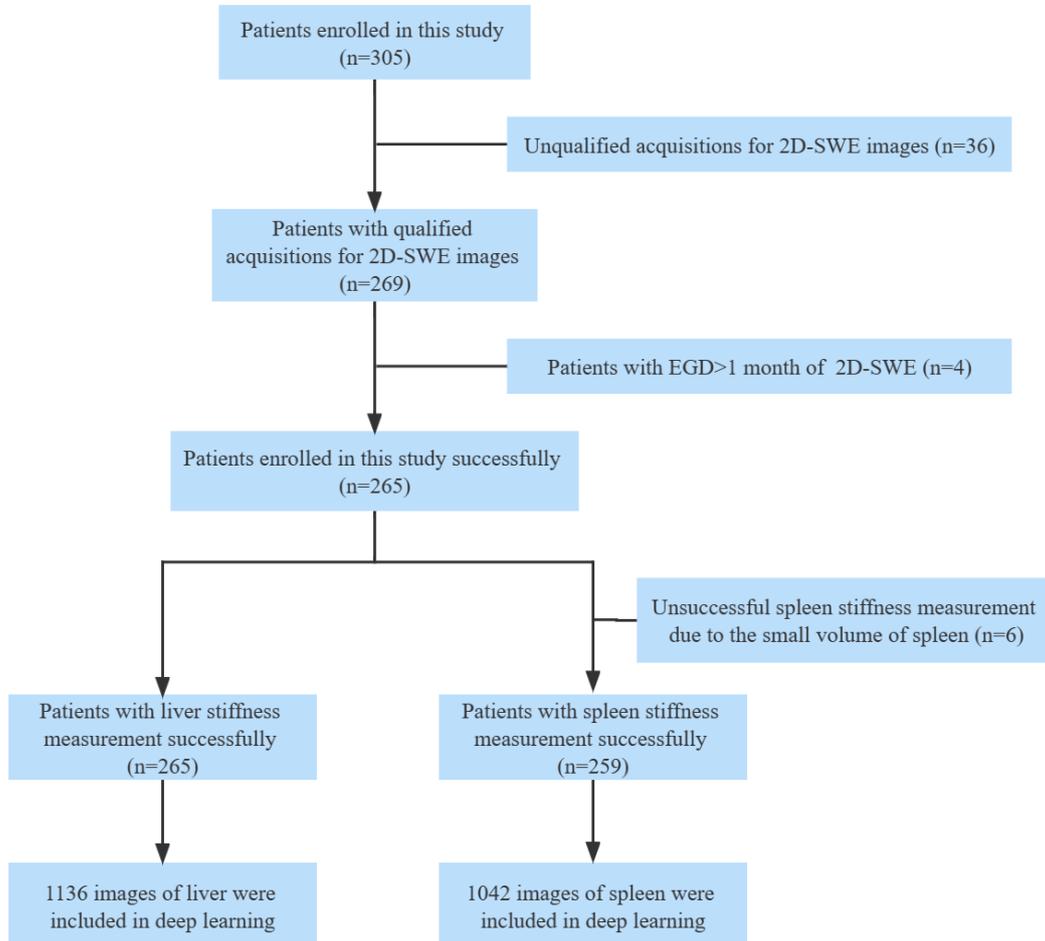

**Figure 2.** Illustration of the 2D-SWE measurement and the DLRP flow chart. (A) Liver stiffness measurement by 2D-SWE. (B) Spleen stiffness measurement by 2D-SWE. (C) A deep learning-based multi-modality classification model was developed to assess the risk of GEV and HRV, based on the liver and spleen features from 2D-SWE images and extracted clinical features from the data of laboratory experiments.

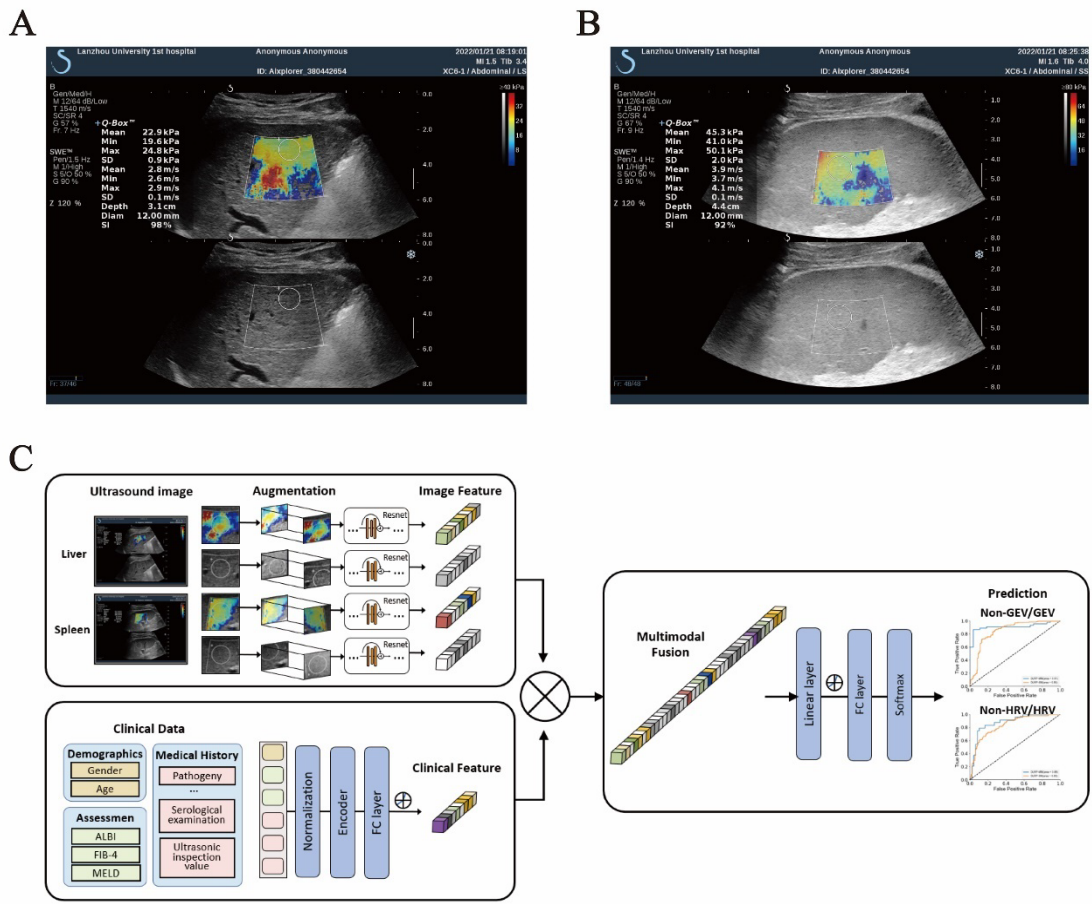

2D-SWE, two-dimensional shear wave elastography; DLRP, multi-modality deep learning risk prediction model; GEV, gastroesophageal varices; HRV, high-risk gastroesophageal varices.

**Figure 3.** Comparison of ROC curves between different methods for classifying whether or not it is GEV/HRV in training and validation cohorts, respectively. (A, C) whether or not it is GEV in training cohorts (A) and validation cohorts (C). (B, D) whether or not it is HRV in training cohorts (B) and validation cohorts (D).

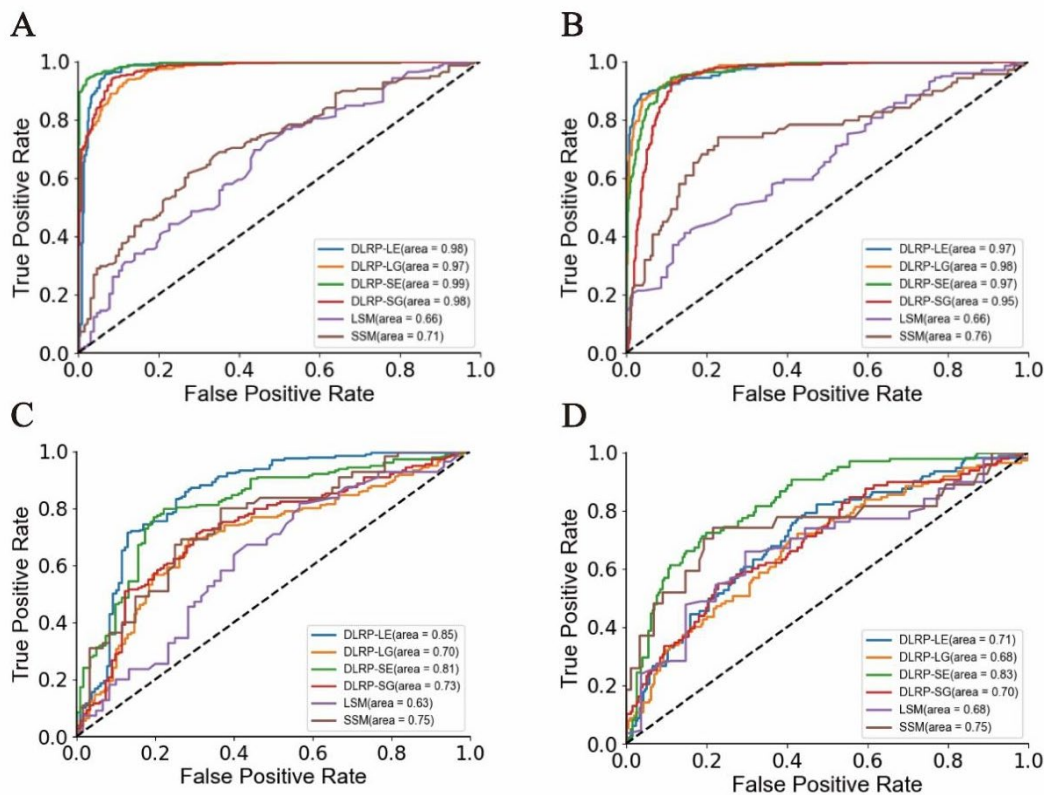

DLRP-LE, deep learning risk prediction model of single-modality with liver elastography images; DLRP-LG, deep learning risk prediction model of single-modality with liver grayscale images; DLRP-SE, deep learning risk prediction model of single-modality with spleen elastography images; DLRP-SG, deep learning risk prediction model of single-modality with spleen grayscale images; LSM, liver stiffness measurement; SSM, spleen stiffness measurement; GEV, gastroesophageal varices; HRV, high-risk gastroesophageal varices. LSM and SSM two comparative experiments using the SVM method.

**Figure 4.** Comparison of ROC curves between multi-modality and single-modality for classifying whether or not it is GEV/HRV in validation cohorts, respectively. (A) whether or not it is GEV in validation cohorts. (B) whether or not it is HRV in validation cohorts.

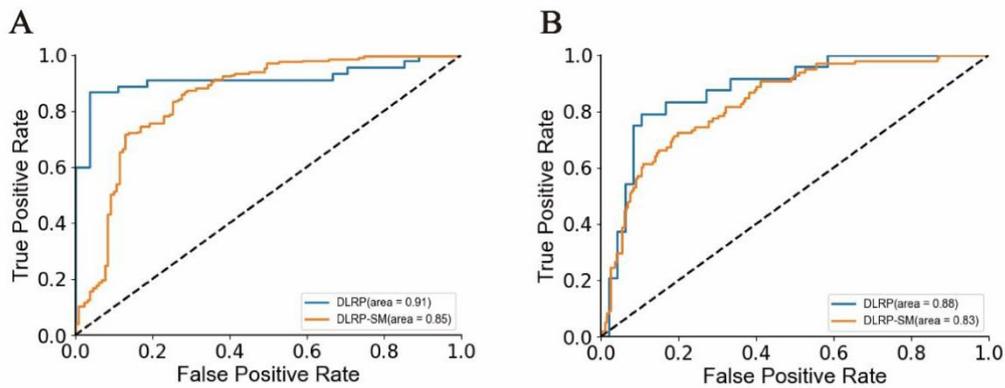

DLRP, multi-modality deep learning risk prediction model; DLRP-SM, deep learning risk prediction model of single-modality; GEV, gastroesophageal varices; HRV, high-risk gastroesophageal varices.

**Figure 5.** Comparison of receiver operating characteristic (ROC) curves between different combinations of hospitals for DLRP in the assessment of GEV and HRV. (A, C) whether or not it is GEV in training cohorts (A) and validation cohorts (C). (B, D) whether or not it is HRV in training cohorts (B) and validation cohorts (D).

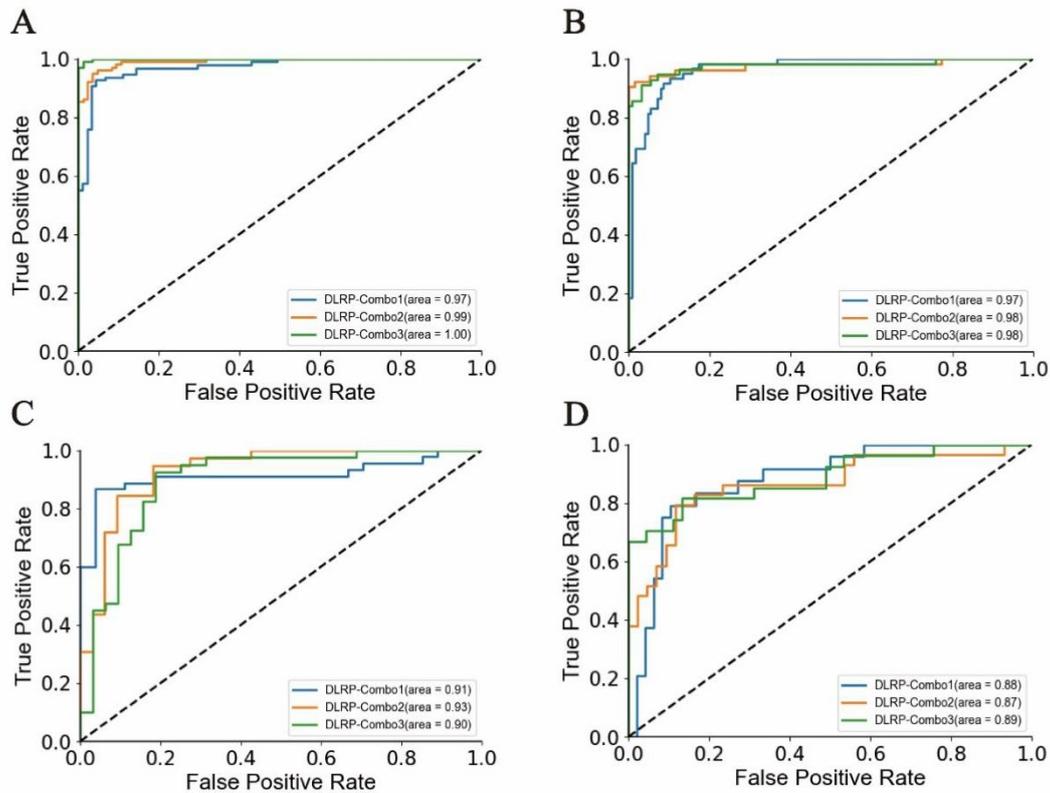

DLRP, multi-modality deep learning risk prediction model; GEV, gastroesophageal varices; HRV, high-risk gastroesophageal varices.

**Table 1.** Baseline characteristics of study cohort.

| | Total (n=265) | HRV cohort (n=83) | Non-HRV cohort (n=182) | p |
|---|---|---|---|---|
| Age (year), median (IQR) | 49.0 (13.0) | 52.0 (12.0) | 48.0 (15.0) | 0.006 |
| Male, n (%) | 177 (66.8) | 57 (68.7) | 120 (65.9) | 0.660 |
| BMI, median (IQR) | 24.0 (4.4) | 24.3 (3.7) | 23.9 (4.5) | 0.544 |
| Etiology, n (%) | | | | - |
|   Hepatitis B infection | 213 (80.4) | 69 (83.1) | 144 (79.1) | |
|   Hepatitis C infection | 16 (6.0) | 5 (6.1) | 11 (6.1) | |
|   Alcoholic liver disease | 7 (2.6) | 2 (2.4) | 5 (2.7) | |
|   Autoimmune | 6 (2.3) | 2 (2.4) | 4 (2.2) | |
|   Primary biliary cirrhosis | 5 (1.9) | 1 (1.2) | 4 (2.2) | |
|   Other | 18 (6.8) | 4 (4.8) | 14 (7.7) | |
| MELD Score, median (IQR) | 9.0 (3.6) | 10.0 (3.4) | 8.4 (3.4) | 0.019 |
| Child-Pugh Class, n (%) | | | | 0.108 |
|   Class A | 219 (82.6) | 64 (77.1) | 155 (85.2) | |
|   Class B | 46 (17.4) | 19 (22.9) | 27 (14.8) | |
| Laboratory, median (IQR) | | | | |
|   PLT ($10^9$/L) | 109.0 (83.5) | 77 (50.0) | 130 (82.5) | <0.001 |
|   AST (U/L) | 39.5 (43.4) | 37.1 (25.4) | 40.5 (56.5) | <0.001 |
|   ALT (U/L) | 38.0 (42.0) | 32.0 (25.0) | 44.5 (69.5) | <0.001 |
|   Alb (g/L) | 39.7 (9.6) | 38.0 (9.1) | 40.6 (9.0) | 0.026 |
|   TBIL (μmol/L) | 22.0 (15.1) | 22.6 (14.9) | 21.0 (13.8) | 0.446 |
| Ultrasound, median (IQR) | | | | |
|   Spleen diameter (mm) | 127.0 (35.8) | 144.0 (33.8) | 120.0 (31.8) | <0.001 |
|   Spleen thickness (mm) | 45.0 (13) | 50.0 (12.8) | 43.0 (9.0) | <0.001 |
| Ultrasound Elastography, median (IQR) | | | | |
|   LSM (kPa) | 14.0 (8.2) | 15.6 (12.1) | 13.1 (7.9) | 0.006 |
|   SSM (kPa) | 32.9 (15.2) | 40.2 (13.3) | 29.5 (12.4) | <0.001 |

IQR, interquartile range; BMI, body index; PLT, platelet count; ALT, alanine aminotransferase; AST, aspartate aminotransferase; GGT, gamma-glutamyl transpeptidase; Alb, albumin; TBIL, total bilirubin; PT, prothrombin time; INR, international normalized ratio; HRV, high-risk gastroesophageal varices; LSM, liver stiffness measurement; SSM, spleen stiffness measurement.

**Table 2.** Diagnostic performance of DLRP-SM for the assessment of GEV in training and validation cohorts.

|  |  | n | AUC | Specificity (%) | Sensitivity (%) | PPV (%) | NPV (%) |
|---|---|---|---|---|---|---|---|
| DLRP-LE | T | 191 | 0.98 | 95.89 | 91.42 | 96.43 | 90.21 |
|  |  |  | (0.97 to 0.98) | (94.57 to 97.12) | (89.87 to 92.90) | (95.19 to 97.50) | (88.23 to 91.98) |
|  | V | 74 | 0.85 | 78.33 | 82.21 | 86.80 | 71.76 |
|  |  |  | (0.83 to 0.87) | (75.66 to 80.89) | (80.64 to 83.91) | (84.76 to 88.68) | (68.67 to 74.81) |
| DLRP-LG | T | 191 | 0.97 \*\*\*^^ | 85.08 | 93.93 | 84.76 | 94.07 |
|  |  |  | (0.97 to 0.98) | (83.27 to 86.95) | (92.45 to 95.36) | (82.58 to 86.91) | (92.5 to 95.52) |
|  | V | 74 | 0.70 \*\*^^^ | 55.11 | 77.63 | 59.90 | 74.05 |
|  |  |  | (0.68 to 0.73) | (51.12 to 57.02) | (75.43 to 79.76) | (57.12 to 62.53) | (71.05 to 76.94) |
| DLRP-SE | T | 187 | 0.99 | 96.66 | 93.73 | 96.77 | 93.53 |
|  |  |  | (0.99 to 0.99) | (95.56 to 97.77) | (92.43 to 95.10) | (95.70 to 97.90) | (92.10 to 95.00) |
|  | V | 72 | 0.81 | 71.88 | 82.56 | 79.78 | 75.41 |
|  |  |  | (0.79 to 0.83) | (69.46 to 74.24) | (80.77 to 84.41) | (77.49 to 81.96) | (72.32 to 78.35) |
| DLRP-SG | T | 187 | 0.98 \*\* | 91.49 | 92.62 | 91.37 | 92.72 |
|  |  |  | (0.97 to 0.98) | (89.92 to 93.10) | (91.07 to 94.10) | (89.60 to 93.10) | (91.10 to 94.20) |
|  | V | 72 | 0.73 \*^ | 62.04 | 77.30 | 70.79 | 69.67 |
|  |  |  | (0.70 to 0.75) | (59.85 to 64.41) | (75.45 to 79.34) | (68.29 to 73.27) | (66.54 to 73.06) |
| LSM | T | 191 | 0.66 \*\*\*^^^ | 55.14 | 67.90 | 39.86 | 79.69 |
|  |  |  | (0.63 to 0.68) | (53.66 to 56.61) | (64.57 to 71.12) | (36.84 to 42.72) | (77.02 to 82.22) |
|  | V | 74 | 0.63 \*\*\*^^^ | 54.12 | 53.33 | 29.09 | 76.67 |
|  |  |  | (0.60 to 0.65) | (52.78 to 55.43) | (49.53 to 57.06) | (26.36 to 32.01) | (74.11 to 79.20) |
| SSM | T | 187 | 0.71 \*\*\*^^ | 63.70 | 67.94 | 64.49 | 67.18 |
|  |  |  | (0.69 to 0.73) | (61.53 to 65.87) | (65.90 to 70.02) | (61.52 to 67.41) | (64.35 to 69.96) |
|  | V | 72 | 0.75 \*\*^^^ | 70.69 | 66.67 | 69.09 | 68.33 |
|  |  |  | (0.73 to 0.77) | (68.60 to 72.81) | (64.44 to 68.84) | (66.11 to 72.07) | (65.48 to 70.95) |

Statistical quantifications were demonstrated with 95% CI, when applicable.

AUC of DLRE-SE was statistically compared with AUC of DLRE-LE, DLRE-LG, DLRE-SG, LSM and SSM, respectively (\*P<0.05; \*\*P<0.01; \*\*\*P<0.001).

AUC of DLRE-LE was statistically compared with AUC of DLRE-LG, DLRE-SE, DLRE-SG, LSM and SSM, respectively (^P<0.05; ^^P<0.01; ^^^P<0.001).

AUC, area under the receiver operating characteristic curve; GEV, gastroesophageal varices; DLRP-SM, deep learning risk prediction model of single-modality; DLRP-LE, DLRP-SM with liver elastography images; DLRP-LG, DLRP-SM with liver grayscale images; DLRP-SE, DLRP-SM with spleen elastography images; DLRP-SG, DLRP-SM with spleen grayscale images; LSM, liver stiffness measurement; SSM, spleen stiffness measurement; n, number of patients; PPV, positive predictive value; NPV, negative predictive value; T, training cohort; V, validation cohort.

**Table 3.** Diagnostic performance of DLRP-SM for the assessment of HRV in training and validation cohorts.

| | | n | AUC | Specificity (%) | Sensitivity (%) | PPV (%) | NPV (%) |
|---|---|---|---|---|---|---|---|
| DLRP-LE | T | 191 | 0.97 | 94.35 | 92.56 | 87.5 | 96.7 |
| | | | (0.96 to 0.98) | (93.20 to 95.44) | (90.60 to 94.50) | (84.83 to 90.05) | (95.83 to 97.66) |
| | V | 74 | 0.71 ** | 73.93 | 54.26 | 45.54 | 80.09 |
| | | | (0.69 to 0.73) | (72.53 to 75.42) | (50.89 to 57.61) | (41.64 to 49.41) | (77.90 to 82.31) |
| DLRP-LG | T | 191 | 0.98 | 94.10 | 89.56 | 87.11 | 95.29 |
| | | | (0.97 to 0.98) | (92.95 to 95.22) | (87.34 to 91.80) | (84.36 to 89.73) | (94.14 to 96.41) |
| | V | 74 | 0.68 *** | 72.96 | 51.58 | 43.75 | 78.70 |
| | | | (0.66 to 0.70) | (71.57 to 74.37) | (48.18 to 54.64) | (40.03 to 47.51) | (76.53 to 80.79) |
| DLRP-SE | T | 187 | 0.97 | 96.82 | 79.34 | 93.48 | 89.06 |
| | | | (0.96 to 0.97) | (95.90 to 97.73) | (76.88 to 81.65) | (91.44 to 95.32) | (87.39 to 90.58) |
| | V | 72 | 0.83 | 84.69 | 65.38 | 69.39 | 82.18 |
| | | | (0.82 to 0.85) | (83.21 to 86.22) | (62.70 to 68.29) | (65.85 to 72.89) | (80.23 to 84.18) |
| DLRP-SG | T | 187 | 0.95 ** | 90.51 | 83.72 | 78.26 | 93.16 |
| | | | (0.94 to 0.96) | (89.26 to 91.83) | (80.90 to 86.41) | (74.96 to 81.58) | (91.81 to 94.49) |
| | V | 72 | 0.70 *** | 73.73 | 56.25 | 36.73 | 86.14 |
| | | | (0.67 to 0.72) | (72.49 to 74.94) | (52.15 to 60.54) | (33.07 to 40.58) | (84.25 to 87.96) |
| LSM | T | 191 | 0.67 *** | 74.06 | - | 0 | 100 |
| | | | (0.64 to 0.69) | | | | |
| | V | 74 | 0.68 *** | 76.52 | - | 0 | 100 |
| | | | (0.65 to 0.70) | | | | |
| SSM | T | 187 | 0.76 *** | 80.35 | 64.86 | 34.78 | 93.40 |
| | | | (0.73 to 0.78) | (79.35 to 81.42) | (59.61 to 70.04) | (30.69 to 39.00) | (92.10 to 94.60) |
| | V | 72 | 0.75 *** | 82.52 | 75.00 | 33.33 | 96.59 |
| | | | (0.72 to 0.78) | (81.65 to 83.50) | (65.44 to 80.39) | (29.21 to 37.74) | (95.69 to 97.45) |

Statistical quantifications were demonstrated with 95% CI, when applicable.
AUC of DLRE-SE was statistically compared with AUC of DLRE-LE, DLRE-LG, DLRE-SG, LSM and SSM, respectively (*P<0.05; **P<0.01; ***P<0.001).
AUC, area under the receiver operating characteristic curve; HRV, high-risk gastroesophageal varices; DLRP-SM, deep learning risk prediction model of single-modality; DLRP-LE, DLRP-SM with liver elastography images; DLRP-LG, DLRP-SM with liver grayscale images; DLRP-SE, DLRP-SM with spleen elastography images; DLRP-SG, DLRP-SM with spleen grayscale images; LSM, liver stiffness measurement; SSM, spleen stiffness measurement; n, number of patients; PPV, positive predictive value; NPV, negative predictive value; T, training cohort; V, validation cohort.

**Table 4.** Performance of DLRP for the assessment of GEV/HRV in training and validation cohorts.

|     |   | n   | AUC | Specificity (%) | Sensitivity (%) | PPV (%) | NPV (%) |
|-----|---|-----|-----|-----------------|-----------------|---------|---------|
| GEV | T | 187 | 0.97*** | 93.33 | 92.78 | 93.75 | 92.31 |
|     |   |     | (0.97 to 0.98) | (91.86 to 94.79) | (91.27 to 94.20) | (92.20 to 95.13) | (90.54 to 93.94) |
|     | V | 72  | 0.91*** | 81.25 | 97.50 | 86.67 | 96.30 |
|     |   |     | (0.90 to 0.93) | (79.10 to 83.33) | (96.61 to 98.34) | (84.80 to 88.48) | (94.93 to 97.60) |
| HRV | T | 187 | 0.97*** | 92.31 | 85.96 | 83.05 | 93.75 |
|     |   |     | (0.96 to 0.97) | (91.08 to 93.55) | (83.49 to 88.42) | (80.03 to 85.90) | (92.47 to 95.03) |
|     | V | 72  | 0.88*** | 88.00 | 81.82 | 75.00 | 91.67 |
|     |   |     | (0.86 to 0.89) | (86.60 to 89.41) | (79.14 to 84.62) | (71.62 to 78.23) | (90.17 to 93.10) |

Statistical quantifications were demonstrated with 95% CI, when applicable.

AUCs of DLRP were statistically compared with deep learning risk prediction model of single-modality in GEV and HRV (*P<0.05; **P<0.01; ***P<0.001).

AUC, area under the receiver operating characteristic curve; DLRP, multi-modality deep learning risk prediction model; GEV, gastroesophageal varices; HRV, high-risk gastroesophageal varices; n, number of patients; PPV, positive predictive value; NPV, negative predictive value; T, training cohort; V, validation cohort.

**Table 5.** Diagnostic robustness performance of DLRP for the assessment of GEV/HRV in training and validation cohorts.

|  |  | n | AUC | Specificity (%) | Sensitivity (%) | PPV (%) | NPV (%) |
|---|---|---|---|---|---|---|---|
| GEV |  |  |  |  |  |  |  |
| Combo 1 | T | 187 | 0.97 | 93.33 | 92.78 | 93.75 | 92.31 |
|  |  |  | (0.97 to 0.98) | (91.86 to 94.79) | (91.27 to 94.20) | (92.20 to 95.13) | (90.54 to 93.94) |
|  | V | 72 | 0.91 | 81.25 | 97.50 | 86.67 | 96.30 |
|  |  |  | (0.90 to 0.93) | (79.10 to 83.33) | (96.61 to 98.34) | (84.80 to 88.48) | (94.93 to 97.60) |
| Combo 2 | T | 187 | 0.99 | 94.19 | 96.04 | 95.10 | 95.29 |
|  |  |  | (0.99 to 1.00) | (92.69 to 95.66) | (94.90 to 97.17) | (93.76 to 96.42) | (93.84 to 96.70) |
|  | V | 72 | 0.93 | 90.00 | 85.71 | 92.31 | 81.82 |
|  |  |  | (0.92 to 0.94) | (88.14 to 91.70) | (83.98 to 87.42) | (90.67 to 93.72) | (79.15 to 84.28) |
| Combo 3 | T | 187 | 1.00 | 98.82 | 98.04 | 99.01 | 97.67 |
|  |  |  | (1.00 to 1.00) | (98.13 to 99.45) | (97.20 to 98.80) | (98.43 to 99.54) | (96.63 to 98.59) |
|  | V | 72 | 0.90 | 86.67 | 85.71 | 90.00 | 81.25 |
|  |  |  | (0.88 to 0.91) | (84.61 to 88.74) | (84.03 to 87.47) | (88.30 to 91.72) | (78.60 to 83.90) |
| HRV |  |  |  |  |  |  |  |
| Combo 1 | T | 187 | 0.97 | 92.31 | 85.96 | 83.05 | 93.75 |
|  |  |  | (0.96 to 0.97) | (91.08 to 93.55) | (83.49 to 88.42) | (80.03 to 85.90) | (92.47 to 95.03) |
|  | V | 72 | 0.88 | 88.00 | 81.82 | 75.00 | 91.67 |
|  |  |  | (0.86 to 0.89) | (86.60 to 89.41) | (79.14 to 84.62) | (71.62 to 78.23) | (90.17 to 93.10) |
| Combo 2 | T | 187 | 0.98 | 96.99 | 94.12 | 92.31 | 97.73 |
|  |  |  | (0.97 to 0.98) | (96.11 to 97.77) | (92.09 to 95.99) | (89.91 to 94.34) | (96.93 to 98.47) |
|  | V | 72 | 0.87 | 84.44 | 81.48 | 75.86 | 88.37 |
|  |  |  | (0.85 to 0.88) | (82.85 to 86.16) | (79.16 to 84.05) | (72.92 to 78.88) | (86.52 to 90.20) |
| Combo 3 | T | 187 | 0.98 | 97.54 | 85.48 | 94.62 | 92.97 |
|  |  |  | (0.97 to 0.98) | (96.72 to 98.33) | (83.04 to 87.79) | (92.76 to 96.38) | (91.59 to 94.25) |
|  | V | 72 | 0.89 | 84.78 | 76.92 | 74.07 | 86.67 |
|  |  |  | (0.87 to 0.90) | (83.19 to 86.42) | (74.37 to 79.53) | (70.93 to 77.20) | (84.80 to 88.48) |

Statistical quantifications were demonstrated with 95% CI, when applicable.

AUCs obtained by three different combinations of patients were statistically compared with each other in each classification and each cohort (*P<0.05; **P<0.01; ***P<0.001).

AUC, area under the receiver operating characteristic curve; DLRP, multi-modality deep learning risk prediction model; GEV, gastroesophageal varices; HRV, high-risk gastroesophageal varices; Combo, combination of patients for training and validation cohorts; n, number of patients; PPV, positive predictive value; NPV, negative predictive value; T, training cohort; V, validation cohort.